# Thermo - mechanical instabilities in friction contact

Valentin L. Popov[1], Andreas Fischersworring-Bunk[2]

[1]Technische Universität Berlin, Dept. of System Dynamics and the Physics of Friction,
10623 Berlin, Germany

[2] MTU AeroEngines AG, Senior Consultant material and damage models, probabilistics,
formerly BMW Group, power train development

The phenomenon of corrugated surfaces is a known technical problem of tribological systems; considerable work has been published in the past on the aspect of rail corrugation of railway systems. Less known is a similar phenomenon observed within the cylinder-piston system of advanced automotive engines using aluminium cylinders. This paper investigates the condition leading to cylinder corrugation in the piston/cylinder system. Material investigations strongly indicate that heat in the contact is playing a major role. Using basic analytical relationships from contact mechanics, the condition required for the onset of such thermo-mechanical instabilities are investigated. Using the concept of a critical velocity it is shown that such instabilities can occur for a realistic set of parameters. A significant technical key factor is the friction coefficient.

## 1  Introduction

The development of instabilities in a tribological system can lead to a permanent deformation pattern (corrugation) of the friction surface and can result in it's functional deterioration. This phenomenon is best known from the rail - wheel system documented in numerous publications [1-6], however it is known to a much lesser extent from automotive internal combustion engines.

Already since the 1970's hypereutectic aluminium-silicon alloys are used in the development of crankcases of automotive internal combustion engines as an alternative material solution. The primary Si precipitation results in a wear resistant surface making the use of grey cast iron liner technology superfluous. This results in a more compact and lightweight design. The excellent thermo-physical properties with high thermal conductivity are especially attractive for high performance engines because of a reduced thermal loading of the liner surface. With the increase of the specific power however the wear resistance is a concern. Already in the early 1990's a wear phenomena was reported specific to high performance engines: Next to the wear marks in the top centre (TC) and bottom centre (BC) crank position of the top piston ring a corrugated surface with a 'washboard pattern' was observed on the liner (see figure 1) in the contact surface between liner and piston rings.

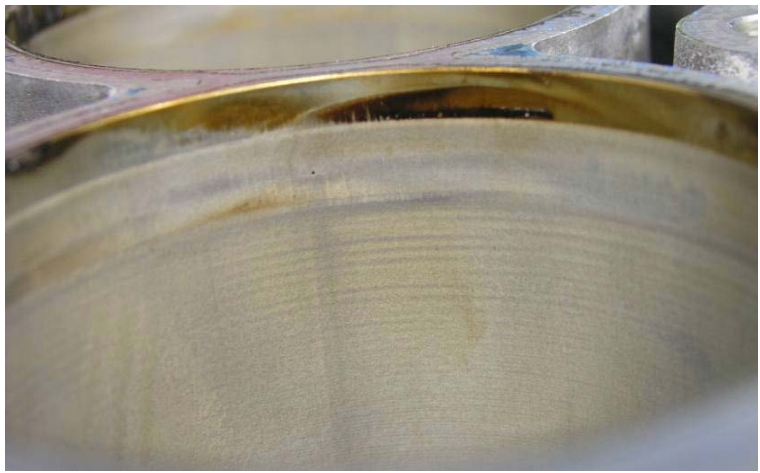

**Figure 1**: View of a corrugated (washboard pattern) cylinder

This wear can develop to a different extent and results in the most detrimental case in a total loss of the engine. In general, the wear is more pronounced in the longitudinal axis of the engine than at the thrust and anti-thrust side. Wear in the range of 1/10 mm can develop even after short engine running times. A particular characteristic is the almost constant wavelength in the wear pattern despite the significant change in the piston velocity. In general, the engineering problem is solved without a thorough understanding of the root cause. The list of the potential influence parameters includes the piston secondary motion, the piston ring design and its manufactured surface condition, the Al-Si cylinder surface condition, and the lubrication. High resolution material microstructure examination using TEM (Transmission Electron Microscopy) of the cylinder reveals that subsurface material is softened by dissolution of the hardening precipitations and also shows a low dislocation density. This change in microstructure can be attributed to the influence of high temperature pointing towards an excessive friction heat. One principal question is related to the topographic regularity of the wear pattern.

This paper tries to explain the washboard wear pattern formation with the help of thermomechanical instabilities. If two elastic bodies in contact are put into relative motion, the interrelation of the released friction energy and the local thermal expansion can result in an instability: areas with higher temperature and hence higher thermal expansion are exposed to higher normal stresses and therefore will heat up even more (see figure 2). The required conditions for the development of such an instability will be reviewed in the following.

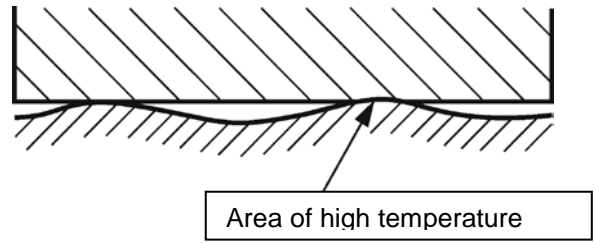

Area of high temperature

**Figure 2**: Areas with higher temperature bulge due to their thermal expansion, resulting in an increased friction energy dissipation and therefore an increase in local temperature. This can lead to an instability and permanent pattern formation.

## 2 Instability problem in a infinite system

We model a system of two bodies, one of which is rigid and has zero-heat conductance. In addition we assume no variation in transverse direction. The boundary between a stable and non-stable condition is defined by a stationary disturbance. Therefore we can assume an equilibrium condition for the elastic body including the effect of the thermal expansion [5,7],

**Fehler! Es ist nicht möglich, durch die Bearbeitung von Feldfunktionen Objekte zu erstellen.** (1)

and the steady heat conductance equation

$$\Delta T = 0 .  \qquad (2)$$

Here $\vec{u}$ denotes the displacement vector, $\nu$ is the Poisson-contraction, T is the deviation of the temperature of it's stationary value far from the surface, $\gamma$ is the thermal coefficient of expansion, and $\Delta = \dfrac{\partial^2}{\partial x^2} + \dfrac{\partial^2}{\partial z^2}$ is the Laplacian-operator. The stress tensor components are given by

$$\sigma_{ik} = -\frac{2}{3}\frac{G(1+\nu)}{(1-2\nu)}\gamma T \delta_{ik} + \frac{2}{3}\frac{G(1+\nu)}{(1-2\nu)}\frac{\partial u_l}{\partial x_l}\delta_{ik} + G\left(\frac{\partial u_i}{\partial x_k} + \frac{\partial u_k}{\partial x_i} - \frac{2}{3}\frac{\partial u_l}{\partial x_l}\delta_{ik}\right). \qquad (3)$$

with shear modulus G. Assuming a rigid surface for the upper body (body 1) we have zero vertical displacement in the elastic body (body 2) (see Figure 3):

$$u_z(z=0) = 0. \qquad (4)$$

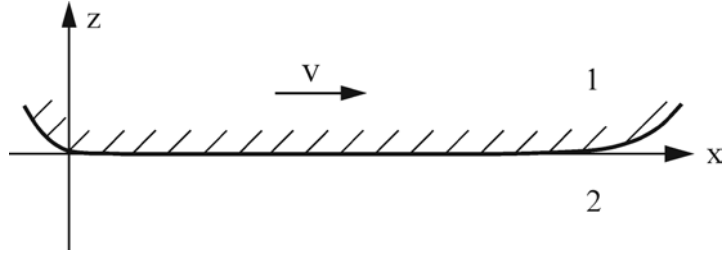

**Figure 3**: A rigid body (1) with zero heat conductivity is in contact with an elastic continua (2). Both bodies are in relative motion to each other with a tangential velocity v.

In view of the situation in an internal combustion engine where we can encounter the thermo-mechanical instability situation, we assume that the coefficient of friction is small and that the normal stress component $\sigma_{zz}$ (e.g. from the piston ring force) is the dominating component. Under these assumptions for the mechanical equilibrium condition the tangential stress component has a negligible contribution:

$$\sigma_{xz}(z=0) = 0. \qquad (5)$$

One solution of equations (1) and (2) subject to the boundary conditions (3) and (4) is given by the solution

$$T = T_0 \cos kx \cdot e^{kz}, \quad \vec{u} = -\frac{\gamma T_0 (1+\nu)}{6(1-\nu)k}\left((-1+kz)\sin kx,\ 0,\ -kz\cos kx\right)\cdot e^{kz}. \qquad (6)$$

(The choice of the term coskx in the solution on the coordinate x means that we are examining the development of a harmonic disturbance. Due to the linearity of the problem an arbitrary disturbance can be described by the superposition of the Fourier coefficients with different wave numbers k).

Under stationary conditions the heat released at the surface must equal the heat flux from the surface (in our model assumption only in the lower body):

$$\kappa \frac{\partial T}{\partial z} = -\mu v \sigma_{zz}(z=0) \qquad (7)$$

with heat conductivity $\kappa$. Therefore the critical value of the wave number is

$$k_c = \frac{v\mu G \gamma (1+\nu)}{3\kappa(1-\nu)}. \qquad (8)$$

For $\nu = 1/3$ the critical wave number is $k_c = \frac{2}{3}\frac{v\mu G\gamma}{\kappa}$. Temperature disturbances with smaller wave numbers than $k_c$ are non-stable.

## 3 Dynamics of an instable disturbance

We will now examine the development of an instable disturbance. For the non-stationary case we can assume that a mechanical equilibrium is achieved and hence equation (1) holds. The steady state heat conductance equation (2) however is replaced by it's unsteady state differential form

$$\frac{\partial T}{\partial t} = \alpha \Delta T \tag{9}$$

with thermal diffusivity α. Again we investigate the evolution of a disturbance which is a periodic function in direction x with wave number k. Solutions of type $T \propto e^{ikx} e^{\lambda z} e^{\omega t}$ fulfil the equation only if the condition

$$\omega = \alpha(\lambda^2 - k^2) \tag{10}$$

holds. A particular solution of equation (1) with regard to the displacement vector $\vec{u}$ is

$$\vec{u} = \frac{\alpha \gamma}{3\omega} \frac{1+\nu}{1-\nu} \nabla T . \tag{11}$$

For the temperature distribution

$$T = T_0 \cos kx \cdot e^{\omega t} e^{\lambda z} \tag{12}$$

we obtain the solution

$$\vec{u}_T = T_0 \frac{\gamma}{3(\lambda^2 - k^2)} \frac{1+\nu}{1-\nu} \left(-k \sin kx,\ 0,\ \lambda \cos kx\right) \cdot e^{\omega t} e^{\lambda z} . \tag{13}$$

Any solution to the homogeneous equation

$$\frac{3}{2} \frac{(1-2\nu)}{(1+\nu)} \Delta \vec{u} + \frac{3}{2} \frac{1}{(1+\nu)} \nabla \operatorname{div} \vec{u} = 0 \tag{14}$$

can be added. These in general have the form

$$\vec{u}_h = \left( (A + Bz)\sin kx,\ 0,\ \left( -\frac{-3B(1-\tfrac{4}{3}\nu) + kA}{k} - Bz \right) \cos kx \right) \cdot e^{kz} . \tag{15}$$

The solution which satisfies the boundary conditions $u_z(z=0)$ and $\sigma_{zx}(z=0) = 0$ is therefore

$$u_x = \frac{T_0}{3} \frac{\gamma(1+\nu)}{\left(\lambda^2 - k^2\right)(1-\nu)} \sin kx \cdot \left[\lambda e^{kz} - k e^{\lambda z}\right] e^{\omega t} \tag{16}$$

$$u_z = \frac{T_0}{3} \frac{\gamma \lambda (1+\nu)}{\left(\lambda^2 - k^2\right)(1-\nu)} \cos kx \left[e^{\lambda z} - e^{kz}\right] e^{\omega t} \tag{17}$$

$$T = T_0 \cos kx \cdot e^{\lambda z} e^{\omega t} . \tag{18}$$

From the heat balance (7) we obtain the equation

$$-3\lambda^2 \kappa (1-\nu) + k\left(-3\lambda \kappa (1-\nu) + 2\mu \nu G \gamma (1+\nu)\right) = 0 \tag{19}$$

from which we can compute the wave number and the magnification constant ω (10) as functions of λ:

$$\frac{k}{k_c} = \tfrac{1}{2} (\lambda/k_c)^2 \frac{1}{1 - \tfrac{1}{2}(\lambda/k_c)} , \tag{20}$$

$$\frac{\omega}{\alpha k_c^2} = (\lambda/k_c)^2 \frac{1 - (\lambda/k_c)}{\left(1 - \tfrac{1}{2}(\lambda/k_c)\right)^2} , \tag{21}$$

or explicitly the magnification constant as a function of the wave number k:

$$\frac{\omega}{\alpha k_c^2} = \frac{2\xi\left(1+\tfrac{1}{2}\xi-\tfrac{1}{2}\sqrt{\xi^2+8\xi}\right)}{\left(1+\tfrac{1}{4}\xi-\tfrac{1}{4}\sqrt{\xi^2+8\xi}\right)} \tag{22}$$

where $\xi = k/k_c$. The dependence is shown in Figure 4.

Within the range of interest for an unstable disturbance it can approximated by

$$\frac{\omega}{\alpha k_c^2} \approx 1{,}5\xi(1-\xi). \tag{23}$$

We again can observe, that all disturbances with a wave number $k < k_c$ are non-stable ($\omega > 0$), meanwhile all disturbances with a larger wave number remain stable.

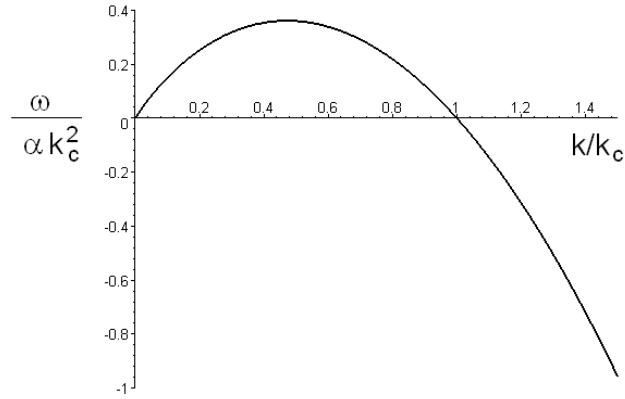

**Figure 4:** Magnification constant $\omega$ as function of the wave number k

## 4 Instability problem in a contact of finite length

Next, we investigate the qualitative condition for a thermo – mechanical instability in a moving contact with *finite* contact length. For a contact of length l, the contact time is $t_1 = l/v$. Under this condition the amplitude of the disturbance of wave vector k will be magnified during the contact time by the factor $\exp(1{,}5\alpha k(k_c - k)l/v)$. If the time till the next contact (half period of the motion) equals $t_2 = L/v$, the disturbance will decay by the factor $\exp(-\alpha k^2 L/v)$ during this time. The condition for instability therefore can be expressed by

$$\exp(1{,}5\alpha k(k-k_c)l/v - \alpha k^2 L/v) = 1. \tag{24}$$

Instability will therefore occur only for disturbances with a wave number smaller than $k_0$, where

$$k_0 = \frac{1{,}5k_c(l/L)}{(1+1{,}5(l/L))} \approx 1{,}5k_c(l/L). \tag{25}$$

This estimate is looses validity if the length of the gliding contact is smaller than the wavelength $\Lambda = 2\pi/k$. The domain of validity is given by the inequality relation

$$2\pi/l < k < 1{,}5k_c(l/L). \tag{26}$$

This relation can only be satisfied if $2\pi/l < 1{,}5k_c(l/L)$ and therefore for velocities larger than

$$v_c \approx \frac{\kappa}{2\pi\mu G\gamma} \frac{L}{l^2}. \qquad (27)$$

However we need to emphasize that this critical velocity must be seen in the context of the used assumptions. Using characteristic material parameters of a hypereutectic aluminium alloy: a shear modulus $G \approx 2.8 \cdot 10^{10}$ Pa, a thermal expansion of $\gamma \approx 1,7 \cdot 10^{-5}$ K$^{-1}$, a heat conductivity of $\kappa \approx 237$ Wm$^{-1}$K$^{-1}$, and using typical engine conditions and dimensions with a friction coefficient in the range $\mu \approx 0,1 - 0,01$, an engine stroke of $L \approx 0.1$ m and a contact length $l \approx 1,5 \cdot 10^{-3}$ m, we obtain a critical velocity of roughly 35-350 *m/s*. We can state that only in the case of relatively high friction coefficients of $\mu \approx 0,1$ the estimated critical velocity is in the range of real piston velocities (for high engine revolutions). Hence the friction coefficient is the most important technical parameter. However the instability can also develop at lower velocities, therefore we will examine the case of point contacts as limiting case the short contacts.

## 5  Estimation of the instability condition for a system in point contact

We will perform a coarse estimate to demonstrate that a thermo - mechanical instability is possible for point contacts as well using the simplified model shown in Figure 5. The point contact can describe e.g. the contact between the piston ring and the cylinder surface.

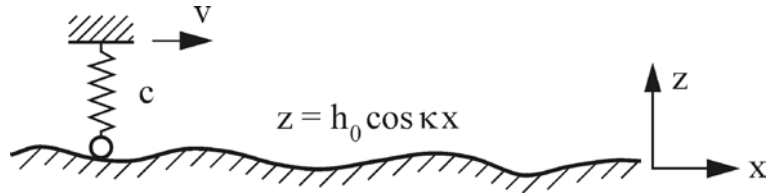

**Figure 5**: Gliding point contact on a wavy surface.

The length of the contact in transverse direction is $L_U$ (in case of the piston ring its circumferential length), and the contact stiffness is c. The contact is moving in x-direction with velocity v. Due to a non-homogeneous heating the surface of the body has a wavy form

$$z = h_0 \cos kx. \qquad (28)$$

Therefore the normal force will depend on the coordinate x. The periodic force component equals

$$\Delta F_N = cz = ch_0 \cos kx. \qquad (29)$$

The resulting change in friction force is computed by

$$\Delta F_R = \mu \Delta F_N = \mu ch_0 \cos kx. \qquad (30)$$

This leads to a heterogeneous heat production on the surface. Due to it's short contact length the heat production will decay during the time $t \approx L/v$ between two subsequent contacts following an exponential law $e^{-k^2\alpha t}$. Therefore only disturbances with wave vectors

$$k \approx (\alpha t)^{-1/2} = \left(\frac{\alpha L}{v}\right)^{-1/2} \qquad (31)$$

will become unstable.

The conditions for instability can be determined as follows. Half the wave length $\Delta x \approx \pi/k$ will be overrun during time $\Delta t \approx \pi/kv$. During that time the heat release equals $\Delta W \approx \mu c h_0 \Delta x = \mu c h_0 \pi/k$. The penetration depth of the heat has the same order of magnitude as half the wave length $\Delta x$. The increase in temperature $\Delta T$ can be estimated from the condition that the heat flux $j \approx \kappa \Delta T / \Delta x = \kappa \Delta T k v / \pi$ equals the friction heat

$$\kappa \Delta T \approx \frac{\mu c h_0 k v}{L_U \pi} \tag{32}$$

per area in square meter and second. The waviness resulting from the thermal expansion equals to $h_0 \approx \frac{\gamma \Delta T \pi}{k}$. With this equation (32) takes the form $\kappa \approx \frac{\mu \gamma c v}{L_U}$. Therefore we obtain the characteristic critical velocity

$$v \approx \frac{\kappa L_U}{\mu \gamma c}. \tag{33}$$

## 6  Summary

We have shown that the interplay of the thermal expansion, the contact mechanics and the frictional heat production in both finite and point contact conditions can result in a thermo-mechanical instability under certain conditions. A system with sufficiently large contact length can be analyzed by analytical means as accomplished in the first two paragraphs. Short contact lengths however will need a numerical analysis. Using a first rough analytical estimate we proved that a thermo-mechanical instability can be caused by a gliding point contact. A thorough numerical analysis is still pending. The thermo-mechanical instabilities and their particular characteristics seem to have simple physical reasons. The characteristic wave number (32) is based in principle on the propagation length of the heat during the half period of the stroke. The most important technical parameter on the critical speed is the coefficient of friction. An instability will only develop for sufficiently high sliding velocities, and results from the balance of heat production and heat conductance.